\documentclass[final,5p,times,twocolumn]{elsarticle}

\biboptions{sort&compress}

\usepackage{color}
\usepackage{amssymb}
\usepackage{amsmath}
\usepackage{mathptmx}
\usepackage{mathrsfs}
\usepackage{footnote}

\journal{Physics Letters B}

\begin{document}

\begin{frontmatter}



\title{Moments of $\phi$ meson spectral functions \\
in vacuum and nuclear matter}


\author[label1]{Philipp Gubler}
\ead{gubler@ectstar.eu}

\author[label1,label2]{Wolfram Weise}
\ead{weise@tum.de}

\address[label1]{ECT*, Villa Tambosi, 38123 Villazzano (Trento), Italy}
\address[label2]{Physik-Department, Technische Universit\"at M\"unchen, 85747 Garching, Germany}

\begin{abstract}
Moments of the $\phi$ meson spectral function in vacuum and in nuclear matter are analyzed, combining a model based on chiral $SU(3)$ effective field theory (with kaonic degrees of freedom) and finite-energy QCD sum rules. 
For the vacuum we show that the spectral density is strongly constrained by a recent accurate measurement of the 
$e^+ e^- \to K^+ K^-$ cross section. 
In nuclear matter the $\phi$ spectrum is modified by interactions of the decay kaons with the surrounding 
nuclear medium, leading to a significant broadening and an asymmetric deformation of the $\phi$ meson peak. 
We demonstrate that both in vacuum and nuclear matter, the first two moments of the 
spectral function are compatible with finite-energy QCD sum rules.  A brief discussion of the next-higher spectral
moment involving strange four-quark condensates is also presented.
\end{abstract}





\end{frontmatter}


\section{\label{Intro} Introduction}
The study of vector mesons ($\rho$, $\omega$ and $\phi$) in nuclear matter has attracted much interest during the last two 
decades \cite{Hayano,Leupold}. In recent years, the $\phi$ meson has particularly come into focus, with dedicated experiments investigating its in-medium properties conducted for instance at KEK \cite{Muto} and at COSY-ANKE  \cite{Polyanskiy}. More detailed measurements are being planned for the future in the E16 experiment at 
J-PARC \cite{Kawama,Aoki}. 
Interpreting the experimental findings requires a thorough theoretical understanding of the modification of the $\phi$ meson spectral function 
at finite density. 

One important issue that needs to be understood is whether and how the modifications of the $\phi$ meson spectral density in nuclear matter 
reflect changes of the non-perturbative QCD vacuum at finite densities. This question has been investigated 
previously in the context 
of QCD sum rules at finite density \cite{Hatsuda,Klingl,Zschocke,Gubler}. 
Using updated input we argue in the present work that the two lowest (zeroth and first) moments are especially suitable for a detailed study of the spectral function with respect to low-dimensional QCD condensates. These lowest moments  
involve only operators up to dimension 4 which are relatively well understood. Condensates of dimension 6 and higher (such as the four-quark condensates) do not yet enter at that stage. 
Furthermore, the ratio of the first over the zeroth moment provides a well defined quantity representing a squared mass averaged over the $\phi$ resonance plus low-energy continuum. This ratio does not depend on details of the spectral function and can in principle be accessed by experimental measurements \cite{Kwon}. 

This article presents a systematic analysis of the lowest two moments of the $\phi$ meson spectral function, using finite-energy QCD sum rules (FESR). To describe the spectral function in vacuum we employ a generalized and improved vector dominance model \cite{Klingl2,Klingl3} and constrain 
its parameters by recent $e^+ e^- \to K^+ K^-$ cross section data \cite{Lees}. 
The changes of this spectrum in nuclear matter are expressed using 
updated kaon-nucleon forward scattering amplitudes, 
with interactions derived from chiral $SU(3)$ 
effective field theory and coupled channels \cite{Ikeda}. 
The resulting spectral functions are then tested for their consistency with FESR. 
Also included is a short digression on higher moments and the strange four-quark condensate. 
A summary and conclusions follow in the final section. 

\section{\label{Vacuum} Spectral moment analysis in vacuum}
\subsection{\label{Phen} The vacuum spectral function}
The starting point is the correlator of the strange quark current, $j_{\mu}(x) = \frac{1}{3} \overline{s}(x) \gamma_{\mu} s(x)$, which couples to the physical $\phi$ meson state: 
\begin{equation}
\Pi_{\mu\nu}(q) = i\displaystyle \int d^{4}x \,\,e^{iqx} \langle \mathrm{T} [j_{\mu}(x) j_{\nu}(0)] \rangle_{\rho}. 
\label{eq:veccorr1}
\end{equation}
$\langle \, \rangle_{\rho}$ stands for the expectation value with respect to 
the ground state of nuclear matter at temperature $T=0$ and with density $\rho$. The vacuum case is realized in the limit $\rho = 0$. 
For a $\phi$ meson at rest it suffices to study the (dimensionless) contracted correlator,  
$
\Pi(q^2) = \frac{1}{3q^2} \Pi^{\mu}_{\mu}(q)
$.
Using an improved vector dominance model \cite{Klingl2}, $\mathrm{Im}\,\Pi(q^2)$ can be 
written as 
\begin{equation}
\mathrm{Im}\Pi(q^2) = \frac{\mathrm{Im}\,\Pi_{\phi}(q^2)}{q^2 g_{\phi}^2} 
\Bigg| \frac{(1-a_{\phi})q^2 - \mathring{m}_{\phi}^2}{q^2 - \mathring{m}_{\phi}^2 - \Pi_{\phi}(q^2)} \Bigg|^2.
\label{eq:veccorr3}
\end{equation}
The self-energy $\Pi_{\phi}(q^2)$ (of dimension $mass^2$) is governed by the coupling of the $\phi$ to $K\overline{K}$ loops and their 
propagation \cite{Klingl2}, either in vacuum or in the nuclear medium.  
The bare mass $\mathring{m}_{\phi}$ and the coupling constant $g_{\phi}$ are determined to 
agree with experimental observations. The 
coupling strength is expected to be of the order of the value determined by 
$SU(3)$ symmetry, $g_{\phi} \simeq -3g/\sqrt{2}$, with $g=6.5$. 
Furthermore, the constant $a_{\phi}$ represents the ratio between the $\phi K \overline{K}$ and $\phi \gamma$ couplings 
and should be close to unity \cite{Klingl,Klingl2}. Here we assume $a_{\phi}=1$ which gives a very good fit to 
the experimental $e^+ e^- \to K^+ K^-$ data as will be shown below. 
The $\phi$ self-energy includes the contributions from charged and neutral kaon loops: 
\begin{equation}
\Pi_{\phi}(q^2) = \Pi_{\phi \to K^+ K^-}(q^2) + \Pi_{\phi \to K^0_L K^0_S}(q^2). 
\end{equation}
For specific expressions of the corresponding loop integrals, see \cite{Klingl2,Klingl3}. 

\begin{figure}
\begin{center}
\includegraphics[width=8cm]{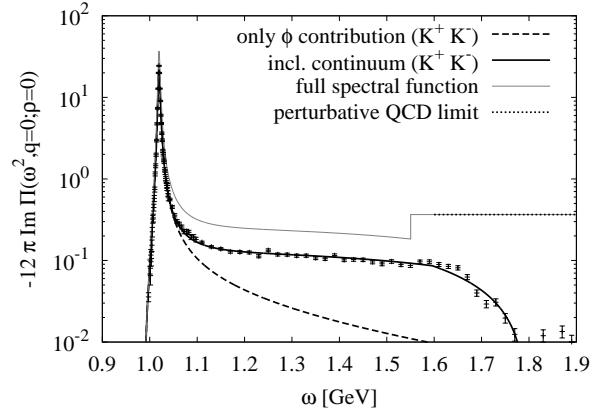}
\end{center}
\vspace{-0.6cm}
\caption{The fitted spectral function $-12 \pi \mathrm{Im} \Pi(\omega^2)$ in vacuum, compared to the experimental data 
for $\sigma(e^+e^- \to K^+ K^-)/\sigma(e^+e^- \to \mu^+ \mu^-)$, adapted from \cite{Lees}. 
The dashed [solid] curve shows the result when only Eq.(\ref{eq:veccorr3}) [both Eqs.\,(\ref{eq:veccorr3}) and (\ref{eq:continuum})] are used for the fit. 
The dotted horizontal line stands for the perturbative QCD limit while the thin gray line represents the full spectral function of Eq.\,(\ref{eq:ansatz}), 
including $K^0\overline{K}^0$ and $K \overline{K} + n \pi$ channels.} 
\label{fig:vac.fit}
\end{figure}

The actual values of $\mathring{m}_{\phi}$ and $g_{\phi}$ are determined by fitting Eq.\,(\ref{eq:veccorr3}) 
to the recent precise measurement of the $e^+ e^- \to K^+ K^-$ cross section
provided by the BABAR Collaboration \cite{Lees}. 
 As in this reaction only the charged kaons are detected, only the corresponding $\phi \rightarrow K^+K^-$ term of  
$\mathrm{Im}\,\Pi_{\phi}(q^2)$ appearing in the numerator of Eq.\,(\ref{eq:veccorr3}) should be kept while 
intermediate charge exchange processes, $K^+K^-\leftrightarrow K^0\overline{K}^0$, are included in the
resummation of the $K\overline{K}$ loops. 
In order to describe the data at energies in the continuum above the $\phi$ meson peak where the 
simple model of Eq.\,(\ref{eq:veccorr3}) cannot be expected to work, we add a  
second order polynomial in $c(q^2) = \sqrt{q^2/q^2_{\mathrm{th}}-1}$, 
for $\sqrt{q^2} > \sqrt{q^2_{\mathrm{th}}}=1040 ~\mathrm{MeV}$: 
\begin{equation}
\mathrm{Im}\Pi^{\mathrm{cont.}}(q^2) = A\,c(q^2) + B\,c^2(q^2),  
\label{eq:continuum}
\end{equation} 
with coefficients $A$ and $B$ fitted to the data. 
This form of the $K^+ K^-$ continuum will be kept both in vacuum and nuclear matter. 
The result of this fit gives $g_{\phi}=0.74 \times (-3g/\sqrt{2}) \simeq -10.2$, 
$\mathring{m}_{\phi} = 797\, \mathrm{MeV}$, $A=-5.94 \times 10^{-3}$ and $B=3.61 \times 10^{-3}$. 
The respective curve is shown in Fig.\,\ref{fig:vac.fit} together with the experimental data. 
As demonstrated in this figure, the parametrizations (\ref{eq:veccorr3},\ref{eq:continuum}) 
give an accurate description of the data up to about $\sqrt{q^2} = \omega\simeq 1.6~\mathrm{GeV}$, above  
which the experimental points are seen to drop rapidly. 
This drop is parametrized by a simple linear curve fitted to the data points in this region. 

Additional channels beyond $e^+ e^- \to K^+ K^-$, such as $K^0\overline{K}^0$ and $K \overline{K} + n \pi$ final 
states, are less well established by empirical data. We include them schematically in the thin solid line shown in Fig.\,\ref{fig:vac.fit}. 

\subsection{\label{Sumrulesvac} Finite-energy sum rules}
In the deep-Euclidean limit ($Q^2 = -q^2 \to \infty$)
the correlator (\ref{eq:veccorr1}) can be expressed with the help 
of the operator product expansion (OPE). The following expansion holds in the vacuum: 
 \begin{equation}
9\,\Pi(q^2=-Q^2) =  -c_0 \log \Big(\frac{Q^2}{\mu^2}\Big) + \frac{c_2}{Q^2} + \frac{c_4}{Q^4} + \frac{c_6}{Q^6} +  \dots.   
\label{eq:ope1}
\end{equation}
For the coefficients $c_i$ one finds\footnote{The $\lambda_a$ in $c_6$ denote Gell-Mann SU(3) color matrices.}
\begin{align} 
c_0 =& \frac{1}{4 \pi^2}\Big(1 + \frac{\alpha_s}{\pi} \Big), \hspace{1cm}
c_2 = -\frac{3 m_s^2}{2 \pi^2}, \label{eq:operesult1} \\
c_4 =& \frac{1}{12} \Big \langle \frac{\alpha_s}{\pi} G^2 \Big \rangle + 2m_s \langle \overline{s} s \rangle, \label{eq:c4} \\
c_6 =& -2 \pi \alpha_s \Big[ \langle (\overline{s}\,\gamma_{\mu} \gamma_5 \,\lambda^a\,s)^2 \rangle  + \frac{2}{9} 
         \langle (\overline{s}\,\gamma_{\mu} \,\lambda^a \,s) \sum_{q=u,d,s} (\overline{q}\,\gamma_{\mu} \,\lambda^a \,q)  \rangle \Big] \nonumber \\
         &+ \frac{m^2_s}{3}\Big[{1\over 3} \Big\langle \frac{\alpha_s}{\pi} G^2 \Big \rangle -8 m_s \langle \overline{s} s \rangle\Big]~. 
\label{eq:operesult3}
\end{align}
Higher order terms in $\alpha_s$ and $m_s$ have also been computed \cite{Gubler}. Here we 
keep only the most important contributions, sufficient for the purposes of the present work. 

Using the once subtracted dispersion relation 
\begin{equation}
\Pi(q^2) = \Pi(0) + \frac{q^2}{\pi}  \displaystyle \int_0^{\infty} ds \frac{\mathrm{Im}\Pi(s)}{s(s - q^2 -i\epsilon)},  
\label{eq:displ}
\end{equation}
and applying the Borel transformation, one derives the sum rule: 
\begin{equation}
\frac{1}{M^2} \int_0^{\infty} ds \, R(s) \, e^{-s/M^2} = c_0 + \frac{c_2}{M^2} + \frac{c_4}{M^4} + \frac{c_6}{2M^6} +  \dots
\label{eq:sumrule}
\end{equation}
with the spectral function 
\begin{equation}
R(s) = -\frac{9}{\pi} \mathrm{Im}\, \Pi(s).
\end{equation}
At large $s$ this spectral function approaches its perturbative QCD limit, so the following ansatz 
is introduced: 
\begin{equation}
R(s) = R_{\phi}(s)\, \Theta(s_0 - s) + R_{\mathrm{c}}(s) \,\Theta(s - s_0), 
\label{eq:ansatz}
\end{equation}
with $R_{\mathrm{c}}(s) = c_0$, and $s_0$ represents a scale that delineates the low-energy and high-energy parts of the spectrum. 
Substituting this into Eq.\,(\ref{eq:sumrule}) and expanding the left-hand side in inverse powers of $M^2$, 
one derives the finite-energy sum rules: 
\begin{align}
\int_0^{s_0} ds \, R_{\phi}(s) &= c_0 \,s_0 + c_2, \label{eq:sm1} \\
\int_0^{s_0} ds \, s \, R_{\phi}(s) &= \frac{c_0}{2} s_0^2 - c_4, \label{eq:sm2} \\
\int_0^{s_0} ds \, s^2 R_{\phi}(s) &= \frac{c_0}{3} s_0^3 + c_6. \label{eq:sm3}
\end{align}
In the following we focus on the first two moments (\ref{eq:sm1},\ref{eq:sm2}) for which the respective Wilson 
coefficients are determined with sufficient accuracy. 
\begin{figure*}
\begin{center}
\includegraphics[width=7cm]{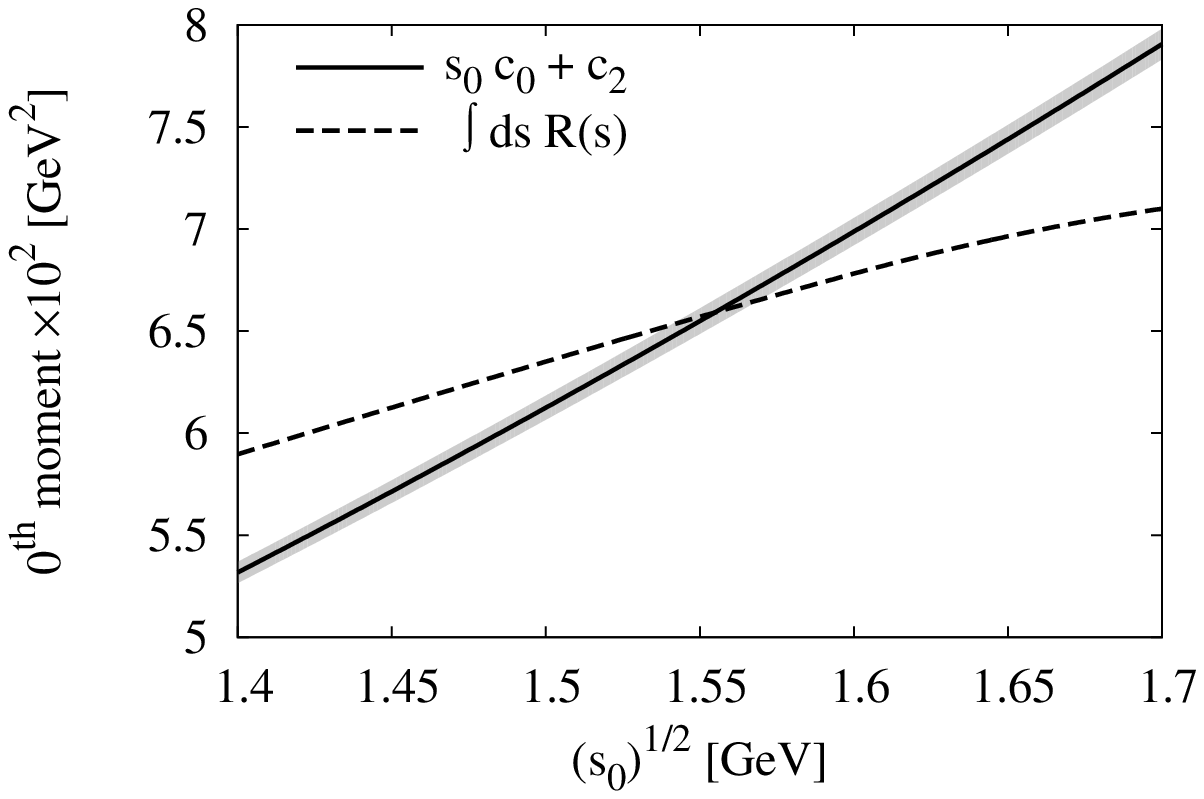}
\includegraphics[width=7cm]{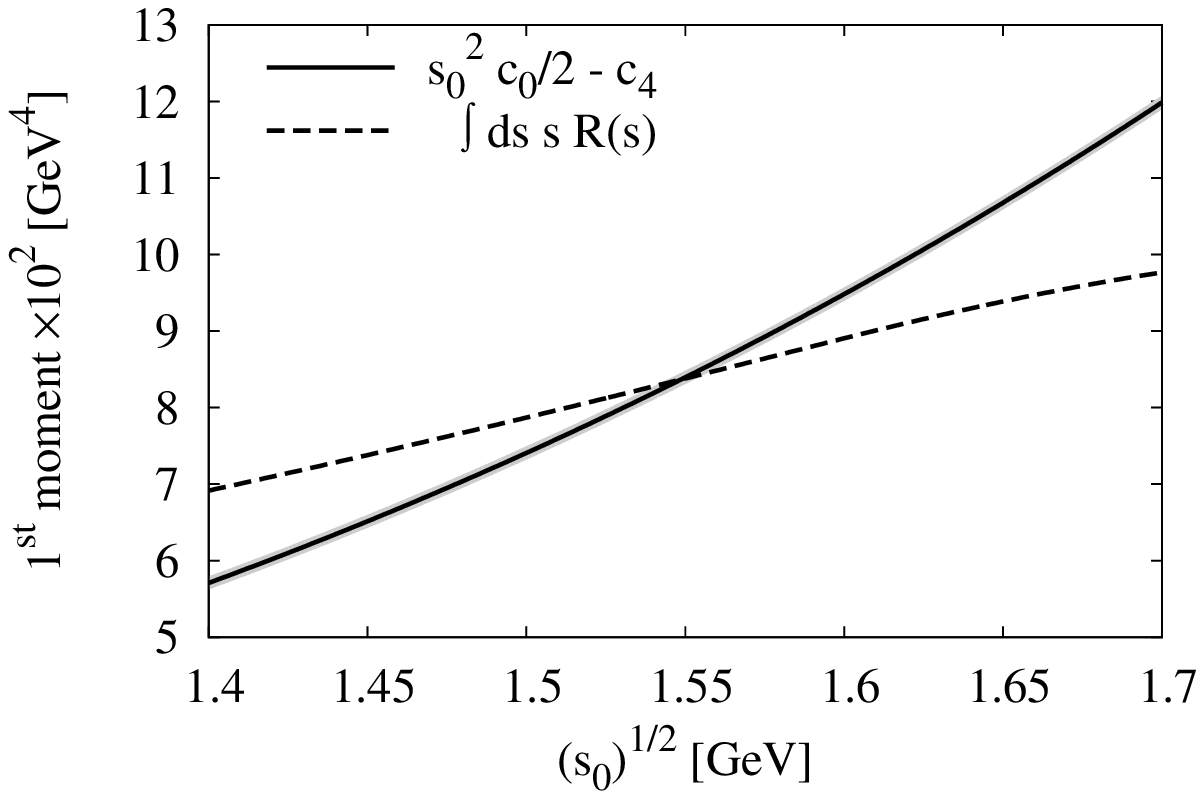}
\end{center}
\vspace{-0.8cm}
\caption{The left-hand and right-hand sides of Eqs.\,(\ref{eq:sm1}) and (\ref{eq:sm2}) as functions of $\sqrt{s_0}$. The error bands (printed in gray) are extracted from the uncertainties of the parameters given in Table \,\ref{tab:parameters}.}
\label{fig:vac.moments}
\end{figure*}

\subsection{\label{Momentanalysisvac} Matching spectral function and sum rules}
In the next step the moments of the spectral function (\ref{eq:veccorr3}) are analyzed using the sum rules (\ref{eq:sm1},\ref{eq:sm2}) and following Refs.\,\cite{Marco,Kwon}. To do this we substitute Eq.\,(\ref{eq:veccorr3}) into the left-hand sides of Eqs.\,(\ref{eq:sm1}) and (\ref{eq:sm2}) and solve them individually for $s_0$. 

In Fig.\,\ref{fig:vac.moments} we compare the left- and right-hand sides of Eqs.\,(\ref{eq:sm1}) and (\ref{eq:sm2}). The parameters entering the right-hand sides are listed in Table \ref{tab:parameters}. 
\begin{table}
\renewcommand{\arraystretch}{1.5}
\setlength{\tabcolsep}{10pt}
\begin{center}
\caption{Parameter values and their uncertainties used for the vacuum QCD sum rule analysis.} 
\label{tab:parameters}
\begin{tabular}{lc}  
\hline 
$\alpha_s(2~\mathrm{GeV})$ & $0.31\pm0.01$ \cite{Olive} \\
$m_s\,(2~\mathrm{GeV})$ & $95\pm5\,\,\mathrm{MeV}$ \cite{Olive} \\
$\langle \overline{s} s \rangle\,(2~\mathrm{GeV})$ & $(-290\pm15\,\,\mathrm{MeV})^3$ \cite{McNeile} \\
$\big \langle \frac{\alpha_s}{\pi} G^2 \big \rangle$ & $0.012 \pm 0.004\,\,\mathrm{GeV}^4$ \cite{Colangelo} \\
\hline
\end{tabular}
\end{center}
\end{table}
We find $\sqrt{s_0} = 1.55\pm0.02~\mathrm{GeV}$ from the zeroth and 
$\sqrt{s_0} = 1.55\pm0.01~\mathrm{GeV}$ from the first moment. One thus observes a remarkable degree
of consistency between these two moments concerning the delineation scale $s_0$ at which perturbative
QCD takes over. 

A characteristic quantity that reflects the $\phi$ mass together with some of the emerging continuum above the resonance can be 
prepared by taking the ratio of the first and zeroth moments, integrated up to a suitably chosen scale $\overline{s}$  with $m_\phi^2 < \overline{s} < s_0$. 
This ratio represents a squared mass averaged over the spectrum $R_{\phi}(s \leq \overline{s})$. 
Choosing $\sqrt{\overline{s}} = 1.2$ GeV turns out to be convenient for this purpose  because the vacuum and in-medium spectral functions 
$R_\phi(s)$ become identical (and constant) above that scale as will be shown. With this $\overline{s} = (1.2$ GeV$)^2$ one finds
\begin{equation}
\overline{m}_{\phi} = \sqrt{\frac{\int_0^{\overline{s}}ds \, s \, R_{\phi}(s)}{\int_0^{\overline{s}}ds \, R_{\phi}(s)}} 
 \simeq 1038 ~\mathrm{MeV}.  
 \label{eq:ratio.vacuum}
\end{equation}
The presence of the continuum above the $\phi$ meson resonance in 
Fig.\,\ref{fig:vac.fit} implies that $\overline{m}_{\phi}$ 
is slightly larger than the physical $\phi$ mass, $m_{\phi}=1019~\mathrm{MeV}$. 

\section{\label{Nuclmatter} Moment analysis in nuclear matter}
Next we extend the FESR analysis to the $\phi$ meson in nuclear matter. 
Working at linear order in the density $\rho$, the in-medium $\phi$ meson self-energy can be 
written: 
\begin{equation}
\Pi_{\phi}(\omega^2) = \Pi_{\phi}^{\mathrm{vac}}(\omega^2) - \rho \mathcal{T}_{\phi N} (\omega), 
\label{eq:rhoT}
\end{equation}
where $\mathcal{T}_{\phi N} (\omega)$ is the free forward $\phi$-nucleon scattering amplitude 
which in turn depends on the interaction between kaon (antikaon) and nucleon \cite{Klingl3}. 
Studies beyond linear order in the density $\rho$ have been performed in Refs. \cite{LutzKorpa,Cabrera,Cieply}.
In the present work we stay at leading order in $\rho$ for reasons of consistency with the sum rule analysis
which is also restricted to in-medium quark and gluon condensates linear in the density. 
In Eq.\,(\ref{eq:rhoT}) the implicit assumption is made that the form, Eq.\,(\ref{eq:veccorr3}), of the vacuum 
self-energy remain unchanged in the medium: a possible additional density dependence of the strange vector-current 
coupling to the $\phi$ meson channel is neglected. We have examined the validity of this assumption by 
introducing an extra factor, $1 - a \rho/\rho_0$, in Eq.\,(\ref{eq:veccorr3}). 
Demanding that the zeroth and first moment sum rules give consistently the same 
scale $s_0$ within errors, we find that this additional factor, at normal nuclear 
matter density, can differ from unity by at most about 2\,\%. Furthermore, we find only a single 
unique solution for which the two $s_0$ values are exactly equal, with the mentioned 
overall factor lying within a 1\,\% range around unity. The above results hold both for 
only S-wave and for S+P-wave $\overline{K}N$ interactions. 
This demonstrates that the sum rules in fact demand that the density dependence of the 
vector current-$\phi$ meson coupling has to be small and that the 
leading density dependence in Eq.\,(\ref{eq:rhoT}) is indeed well represented by the $\rho \mathcal{T}_{\phi N}$ 
term. 

Consider first S-wave $KN$ and $\overline{K}N$ interactions in $\mathcal{T}_{\phi N} (\omega)$. 
For the $\overline{K}N$ system we employ the amplitude derived from $SU(3)$ chiral effective field
theory \cite{Ikeda}, including $\overline{K}N\leftrightarrow\pi\Sigma$ coupled channels and the dynamical generation of the $\Lambda(1405)$. 
This strongly energy-dependent amplitude reproduces all presently 
available scattering data 
together with accurate kaonic hydrogen measurements. 
For the $KN$ channel we follow Ref.\,\cite{Klingl3} and approximate the corresponding (weakly energy-dependent) amplitude by its real part (see also \cite{Waas,Lutz1}). 

P-wave $KN$ and $\overline{K}N$ interactions are incorporated using the spectral function provided in \cite{Klingl3}. 
This spectral function includes the relevant baryon octet and decuplet intermediate states at 
one-loop level. It describes the region around the $\phi$ meson peak up to $\omega \simeq 1.1$ GeV and is then smoothly connected 
to the continuum at higher energies. The $\phi$ spectrum involving pure S-wave $KN$ and $\overline{K}N$ couplings starts at the $K\overline{K}$ threshold, 
located at twice the kaon mass in vacuum and shifted downward in the medium primarily by the attractive $\overline{K}N$ interaction. 
The inclusion of P-wave interactions couples the $\phi$ to the $YK$ continuum with correspondingly lower thresholds, where $Y$ stands for $\Lambda, \Sigma$ and $\Sigma^*$.
 
\begin{figure}
\begin{center}
\includegraphics[width=8cm]{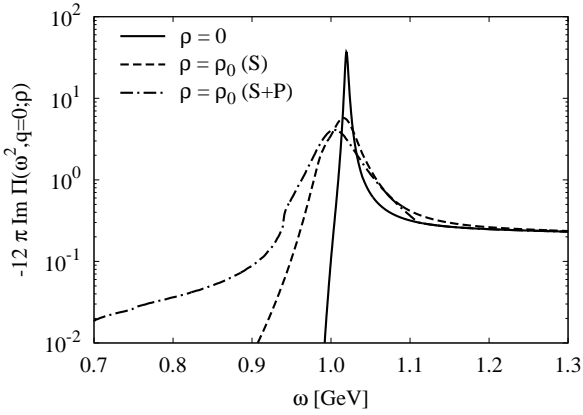}
\end{center}
\vspace{-0.6cm}
\caption{The spectral function $-12 \pi \mathrm{Im} \Pi(\omega^2)$ in vacuum (solid curve) and at normal nuclear matter density, $\rho = \rho_0 = 0.17$ fm$^{-3}$ 
(only S-waves: dashed curve, S- and P-waves: dash-dotted curve).}
\label{fig:vac}
\end{figure}

The resulting spectral functions are shown as solid (vacuum), dashed (only S-wave) and 
dash-dotted (S- and P-wave) curves in Fig.\,\ref{fig:vac}. 

The OPE input for the sum rules is also modified by finite-density effects. 
The vacuum condensates receive the following corrections at leading order in $\rho$: 
\begin{align}
\langle \overline{s} s \rangle_{\rho} \simeq &\; \langle \overline{s} s \rangle + \langle N| \overline{s} s |N \rangle \rho = \langle \overline{s} s \rangle + \frac{\sigma_{sN}}{m_s} \, \rho, \\
\Big \langle \frac{\alpha_s}{\pi} G^2 \Big \rangle_{\rho} \simeq &\; \Big \langle \frac{\alpha_s}{\pi} G^2 \Big \rangle 
+ \Big \langle N \Big| \frac{\alpha_s}{\pi} G^2 \Big|N\Big \rangle \rho \nonumber \\
= &\; \Big \langle \frac{\alpha_s}{\pi} G^2 \Big \rangle - \frac{8}{9}\big(M_N - \sigma_{\pi N} - \sigma_{sN}\big)\,\rho. 
\end{align} 
Here $M_N$ is the physical nucleon mass, $\sigma_{\pi N} = 2m_q \langle N| \overline{q} q |N \rangle$ is the $\pi N$ sigma term and 
$\sigma_{sN} = m_s \langle N| \overline{s} s |N \rangle$ the strangeness sigma term of the nucleon. 
In addition, there is a correction coming from a twist-2 operator given as 
\begin{align}
\mathcal{S}\mathcal{T} \langle N| \bar{s}\, \gamma^{\,\mu} D^{\nu} s |N \rangle = \frac{-i}{2M_N} A^s_2\, \big(p^{\mu} p^{\nu} - \frac{1}{4} M_N^2 \,g^{\mu \nu} \big). 
\end{align} 
The symbols $\mathcal{S}\mathcal{T}$ stand for the operation of making the matrix symmetric and traceless with respect to the Lorentz indices, and $p^{\mu}$ is the four-momentum of the nucleon ($p^2 = M_N^2$). $A^s_2$ is the first moment of the parton distributions of strange quarks in the nucleon\footnote{An additional term related to the first moment 
of the gluon parton distribution \cite{Gubler} is ignored here for simplicity.}
\begin{align}
A^s_2 = 2 \int_0^1 dx~x \big[s(x) + \overline{s}(x) \big]. 
\end{align} 
Altogether, one finds the following in-medium correction to the coefficient $c_4$ of Eq.\,(\ref{eq:c4}): 
\begin{align}
\delta c_4  = \Big(- \frac{2}{27} M_N + \frac{56}{27} \sigma_{sN} + \frac{2}{27} \sigma_{\pi N} + A^s_2 \,M_N \Big) \rho.
\end{align}
The various parameters appearing in this expression are chosen as listed in Table \ref{tab:parameters.matter}. 
\begin{savenotes}
\begin{table}
\renewcommand{\arraystretch}{1.5}
\setlength{\tabcolsep}{10pt}
\begin{center}
\caption{Parameter values and ranges used for the QCD sum rule analysis in nuclear matter. The moment $A^s_2$ of the strange quark 
parton distribution is given at a renormalization scale $\mu = 1$ GeV. (The $25\,\%$ error in $A^s_2$ covers possible uncertainties related 
to an evolution towards $\mu = 2\, \mathrm{GeV}$, the renormalization scale at which the parameters in Table \ref{tab:parameters} 
are determined.)} 
\label{tab:parameters.matter}
\begin{tabular}{lc}  
\hline 
$M_N$ & $940\,\,\mathrm{MeV}$ \\
$\sigma_{sN}$ & $35\pm35\,\,\mathrm{MeV}$ \cite{Young,Babich,Durr,Horsley,Bali,Semke,Freeman,Shanahan,Ohki,Alarcon,Engelhardt,Junnarkar,Jung,Gong,Alexandrou,Lutz,Ren} \\
$\sigma_{\pi N}$ & $45\pm7\,\,\mathrm{MeV}$ \cite{Gasser}\footnote{According to recent analysis \cite{Hoferichter,Alarcon2}, $\sigma_{\pi N}$ may increase to a larger value, 
which however does not affect the conclusions of this work.} \\
$A^s_2$ & $0.044 \pm 0.011$ \cite{Martin,Gubler} \\
\hline
\end{tabular}
\end{center}
\end{table}
\end{savenotes}
For the strangeness sigma term, $\sigma_{sN}$, no generally accepted value is available at present and we hence take a rather broad range that encompasses almost all recent studies \cite{Young,Babich,Durr,Horsley,Bali,Semke,Freeman,Shanahan,Ohki,Alarcon,Engelhardt,Junnarkar,Jung,Gong,Alexandrou,Lutz,Ren}, 
including direct lattice QCD computations and chiral extrapolations of available lattice data. 

\begin{figure*}
\begin{center}
\includegraphics[width=7cm]{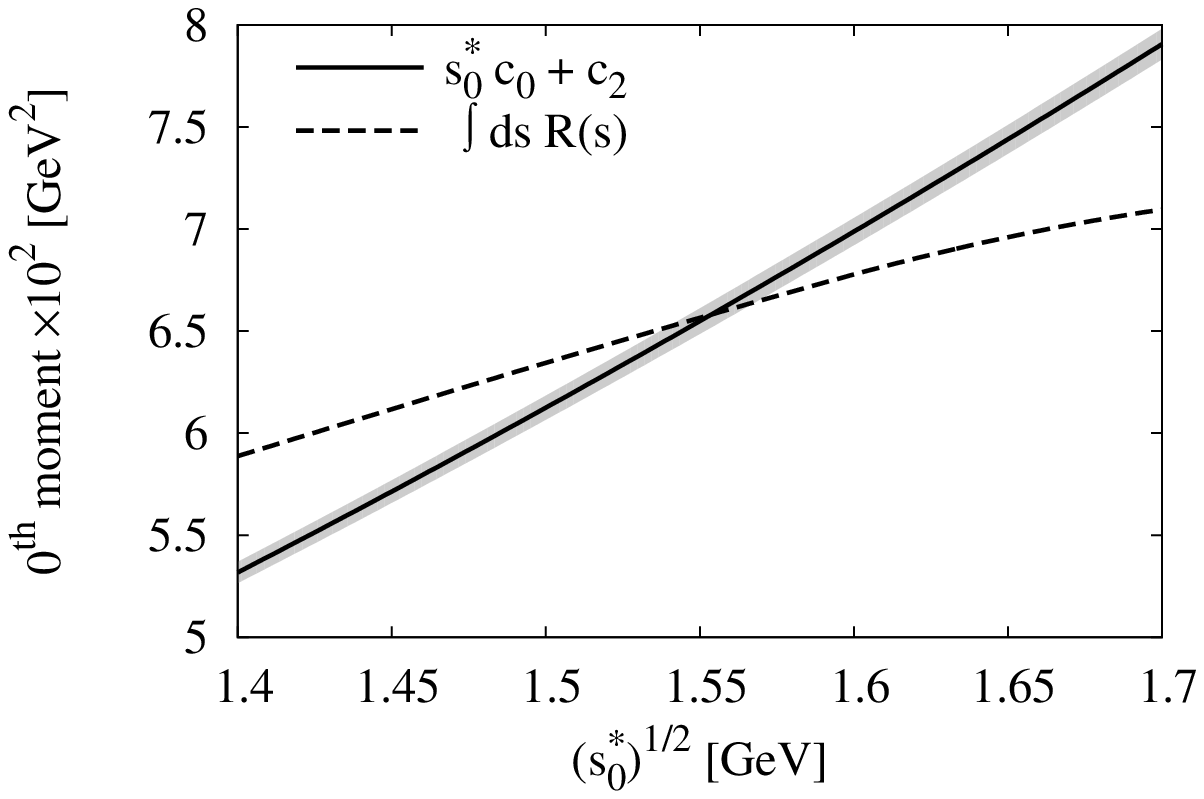}
\includegraphics[width=7cm]{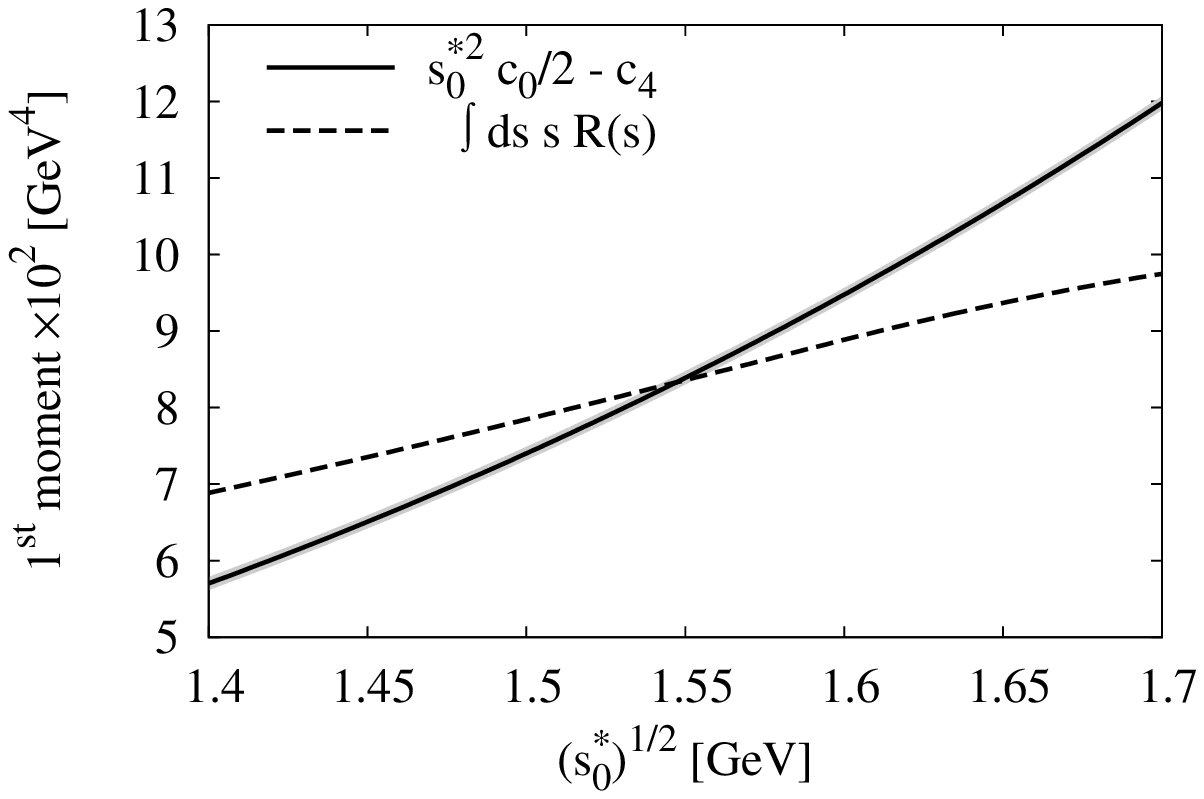}
\end{center}
\vspace{-0.8cm}
\caption{The left- and right-hand sides of the nuclear matter versions of Eqs.(\ref{eq:sm1}) and (\ref{eq:sm2}) as functions of $\sqrt{s_0^{\ast}}$ at normal nuclear matter density, $\rho = \rho_0 = 0.17$ fm$^{-3}$.}
\label{fig:matter.moments}
\end{figure*}

With these inputs, the analysis of in-medium spectral moments is now carried out in the same way as in the previous section. 
As a first step, consider only S-wave kaon- and antikaon-nucleon interactions, using the dashed curve in Fig.\,\ref{fig:vac} as input for the analysis. 
We solve again Eqs.\,(\ref{eq:sm1}) and (\ref{eq:sm2}) for the in-medium delineation scale between low-energy and perturbative QCD regions, now denoted $s_0^{\ast}$ in order to distinguish it  
from the vacuum value $s_0$. The left- and right-hand sides of the equations are shown in Fig.\,\ref{fig:matter.moments} as functions of $\sqrt{s_0^{\ast}}$. 
At normal nuclear matter density, $\rho = \rho_0 = 0.17$ fm$^{-3}$, one finds
$\sqrt{s_0^{\ast}} = 1.55\pm0.02\,\mathrm{GeV}$ from the zeroth moment, and 
$\sqrt{s_0^{\ast}} = 1.55\pm0.01\,\mathrm{GeV}$ from the first moment. At this point both in-medium scales are again consistent with each other and show no change in comparison with the corresponding vacuum values. 

Setting once again $\overline{s} = (1.2\,\mathrm{GeV})^2$ as in the vacuum case, the ratio of first to zeroth moment is then evaluated as 
\begin{equation}
\overline{m}_{\phi}^{\ast}(\rho=\rho_0) = \sqrt{\frac{\int_0^{\overline{s}}ds \,s \,R_{\phi}(s)}{\int_0^{\overline{s}}ds \,R_{\phi}(s)}} 
 \simeq 1035 ~\mathrm{MeV}.  
\end{equation}
Comparing this result with Eq.\,(\ref{eq:ratio.vacuum})
it is seen that the averaged mass $\overline{m}_{\phi}^{\ast}$ at $\rho = \rho_0$ does not deviate from the vacuum value apart from a marginal downward shift. 
In contrast, the resonant $\phi$ meson peak experiences a significant broadening, with a width of $\Gamma_{\phi} \simeq 24\,\mathrm{MeV}$ at normal nuclear matter density. 
The width is determined by the imaginary part of the self-energy at the resonance maximum. 

Effects of P-wave antikaon-nucleon interactions are studied employing the dash-dotted curve of Fig.\ref{fig:vac}. 
The numerical analysis is performed as in the previous paragraphs and with the same OPE input. 
For the scale parameter $s_0^{\ast}$ at $\rho = \rho_0$, we extract $\sqrt{s_0^{\ast}} = 1.52\pm0.02\,\mathrm{GeV}$ from 
the zeroth moment, and $\sqrt{s_0^{\ast}} = 1.52\pm0.01\,\mathrm{GeV}$ from the first moment. 
The behavior of the 
left- and right-hand sides of Eqs.\,(\ref{eq:sm1}) and (\ref{eq:sm2}) is similar to the S-wave case and we do 
not show it here. 

Finally, setting $\overline{s} = (1.2\,\mathrm{GeV})^2$ as before, the ratio of first to zeroth moments at $\rho = \rho_0$ including both S- and P-wave interactions is computed as:  
\begin{equation}
\overline{m}_{\phi}^{\ast}(\rho=\rho_0) = \sqrt{\frac{\int_0^{\overline{s}}ds\, s \,R_{\phi}(s)}{\int_0^{\overline{s}}ds\,R_{\phi}(s)}} 
 \simeq 1022~\mathrm{MeV}.  
\end{equation}
Comparing this result with the vacuum value $\overline{m}_{\phi}$ of Eq.(\ref{eq:ratio.vacuum}), the averaged mass now experiences a modest downward shift by about $16\,\mathrm{MeV}$, more than for the pure S-wave case. This difference is explained by both 
a small shift of the $\phi$ resonance peak and the pronounced low-energy continuum in the spectral function, caused by the opening of kaon-hyperon channels in the presence of P-wave interactions (see Fig.\,\ref{fig:vac}). 
The broadening of the in-medium $\phi$ resonance is further increased significantly, reaching $\Gamma_{\phi}(\rho =\rho_0) \simeq 45\,\mathrm{MeV}$, an order of magnitude larger than the vacuum width ($\Gamma_{\phi}(0) = 4.3\,\mathrm{MeV}$). 

\section{\label{Fourquark} Strange four-quark condensates in vacuum and nuclear matter}
So far our FESR analysis has been restricted to the lowest two moments of $R_\phi(s)$, Eqs.(\ref{eq:sm1},\ref{eq:sm2}). 
Let us now briefly turn to the next-higher moment and the respective sum rule, Eq.(\ref{eq:sm3}), and study its 
implications for the vacuum and in-medium four-quark condensates. 
First, we introduce the commonly used parametrization based on the factorization hypothesis: 
\begin{align}
\langle (\overline{s}\,\gamma_{\mu} \gamma_5,\lambda^a\,s)^2 \rangle  +& \frac{2}{9} 
\langle (\overline{s}\,\gamma_{\mu} \,\lambda^a\, s) \sum_{q=u,d,s} (\overline{q}\,\gamma_{\mu} \,\lambda^a\, q)  \rangle \nonumber \\
= & 
\frac{112}{81}\,\kappa_0 \,\langle \overline{s} s \rangle^2. 
\label{eq:fourquark1}
\end{align}
Exact vacuum saturation corresponds to $\kappa_0 = 1$. Any deviation from this value 
signals the degree of violation of the factorization assumption. 

At finite density, the right-hand side of Eq.(\ref{eq:fourquark1}) is changed to 
\begin{equation}
\frac{112}{81} \kappa_N(\rho)  \Big( \langle \overline{s} s \rangle^2 + 2 \frac{\sigma_{sN}}{m_s} \, \langle \overline{s} s \rangle \, \rho \Big), 
\label{eq:fourquark2}
\end{equation}
where $\kappa_N(\rho)$ can now in general depend on the density $\rho$ and the in-medium strange quark condensate has been expanded to leading order in $\rho$. 
Let us furthermore note that as long as $\kappa_0$ or $\kappa_N$ take values of order 1 and larger, the four-quark condensate term 
in Eq.\,(\ref{eq:operesult3}) dominates over the other two terms of higher order in $m_s$. These terms can thus be safely neglected for the purposes of this discussion. 
Further terms that could show up at dimension 6 include the mixed condensate ($m_s \langle \overline{s} g \sigma G s \rangle$) 
and the gluon condensate involving three gluon fields ($\langle g^3 G^3 \rangle $). For the vector correlator, the Wilson coefficients of 
both these condensates are known to vanish at leading order in $\alpha_s$ and are therefore suppressed. 

\begin{figure}
\begin{center}
\includegraphics[width=8cm]{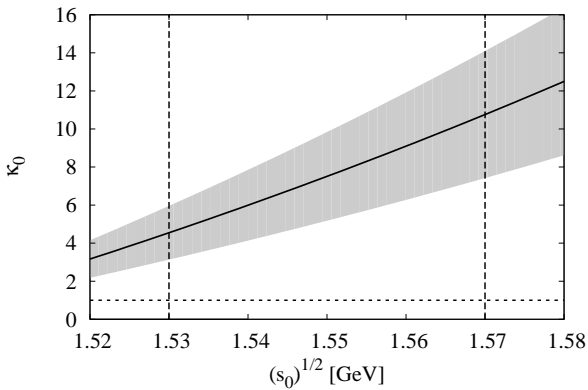}
\end{center}
\vspace{-0.8cm}
\caption{The factorization parameter $\kappa_0$ of the four-quark condensate defined in Eq.\,(\ref{eq:fourquark1}), 
extracted from the second moment sum rule (\ref{eq:sm3}), as a function of the scale $\sqrt{s_0}$ delineating low-energy and PQCD regions. 
The vertical dashed lines show the bounds of $\sqrt{s_0} = 1.55 \pm 0.02$, as obtained from the zeroth moment sum rule. 
The horizontal short-dashed line indicates $\kappa_0=1$, corresponding to the exact factorization hypothesis. 
}
\label{fig:four.quark}
\end{figure}

Next, we combine our spectral function and the threshold parameters obtained from the zeroth and first 
moments with the FESR of Eq.(\ref{eq:sm3}) in order to examine possible constraints for $\kappa_0$ and $\kappa_N$. 
For the vacuum case the result of this investigation is summarized in Fig.~\ref{fig:four.quark}.  
The large uncertainties in $\kappa_0$ 
can be understood from the fact that, compared to the leading order perturbative QCD term proportional to
$s_0^3$, the $c_6$ in Eq.(\ref{eq:sm3}) gives only a small correction to the sum rule. The strange four-quark
condensate is therefore not so well constrained. A strong dependence of $\kappa_0$ on the 
scale parameter $\sqrt{s_0}$ is observed. For a typical value,  $\sqrt{s_0} = 1.55$ GeV, one finds 
\begin{equation}
\kappa_0 \sim 7 \pm 2. 
\end{equation}
Despite such large uncertainties, it can be concluded at least qualitatively that factorization $(\kappa_0 = 1)$
is not expected to be a good approximation. Repeating this analysis for the in-medium case with S- and P-wave
kaon- and antikaon-nuclear interactions shows a similar tendency for $\kappa_N(\rho)$, indicating that the 
factorization hypothesis appears to be strongly violated in nuclear matter. 
At finite density, a full analysis including additional, potentially large twist-2 and twist-4 terms will however be 
needed to reach a definitive conclusion (see \cite{Gubler2} for a discussion of some of these terms). 
This issues deserve a more 
detailed investigation that we plan to pursue in a forthcoming publication.    


\section{\label{Conclusion} Summary and Conclusions}
This work has been focused on the $\phi$ meson spectral functions in vacuum and in 
nuclear matter using new and updated input for the computation of $\phi \leftrightarrow K\overline{K}$ loops and their interactions with nucleons in the surrounding medium. A test of these spectral functions has been performed in comparison with QCD finite-energy sum rules (FESR). The detailed investigation of FESRs for the zeroth and and first moments of these spectral functions demonstrates consistency with well established QCD operators and condensates involving strange quarks and gluons up to dimension four. This non-trivial conclusion holds both in the vacuum and for in-medium quantities at leading order in the density. Less well determined operators of higher dimension, such as four-quark condensates, appear in conjunction with the second spectral moment and have also been briefly examined and discussed. While the FESR results for this next-higher moment
are subject to relatively large uncertainties, a qualitative conclusion can already be drawn, namely that the 
frequently used factorization (i.e. ground state saturation) hypothesis is not likely to work, either in the vacuum or in the nuclear medium. 

The $\phi$ spectral function in vacuum is constructed combining an improved vector dominance approach with a 
background parametrization that fits the recent high-quality $e^+e^-\rightarrow K^+K^-$ data provided by the BABAR collaboration. The resulting zeroth and first spectral moments are in perfect agreement with the
FESR analysis. In the nuclear medium the $\phi$ resonance experiences strong broadening, mainly from the in-medium interactions that couple the $\phi \rightarrow K\overline{K}$ loop to $K\overline{K}N \rightarrow K$-hyperon continuum channels. Induced by such mechanisms the $\phi$ width increases to about $45\,\mathrm{MeV}$
at normal nuclear matter density $\rho_0$, a factor of ten larger than the vacuum decay width. The in-medium (downward) mass shift of the $\phi$ turns out to be small. At $\rho = \rho_0$ it amounts to less than 2\% of its vacuum mass. This confirms findings of earlier studies based on chiral SU(3) models \cite{Klingl3,Cabrera}. 

In this work, we have restricted ourselves to the case of a $\phi$ meson at rest with respect to the nuclear medium. 
Some caution is therefore needed when comparing our results to experiments, in which the measured spectrum always 
involves finite momentum contributions \cite{Muto}. 
Effects of non-zero momenta have so far only been investigated in the sum rule approach of \cite{Lee}. 
More detailed systematic studies are required to fully understand the finite-momentum effects, 
especially in view of the E-16 experiment at J-PARC, where $\phi$ meson spectra at a sequence of 
different momenta are planned to be measured \cite{Kawama,Aoki}. 

Further quantitative improvements of the present analysis will include, for instance, implementing a more realistic treatment of the onset of the continuum. In the present work the delineation between the low-energy hadronic
region and the QCD continuum has been introduced by a simple step-function at a scale $s=s_0$ 
which turned out to emerge consistently at $\sqrt{s_0}\simeq 1.5$ GeV. A softer transition represented by 
a ramp function \cite{Kwon} should make sure that conclusions do not depend on details of this
ramping into the continuum. An extended publication, including a full discussion of such refinements,
is in preparation. 

Finally let us point out implications of the present study for future measurements of the in-medium $\phi$ meson spectral function, in particular with reference to the E16 experiment planned to be performed at J-PARC. A characteristic feature of the in-medium spectra shown in Fig. \ref{fig:vac} is their non-symmetric behavior around the $\phi$ meson peak, with an enhancement of strength in the low-energy sub-resonance region caused by strong broadening and the opening of new decay channels. Such behavior should be taken into account when analyzing the experimental spectra. In addition we once more emphasize the usefulness of the two lowest spectral moments as they provide a direct link to the most relevant low-dimensional QCD condensates and their respective changes in nuclear matter. Precise measurements 
of these moments with nuclear targets would make it possible to constrain the behavior of the gluon and strange quark condensates at finite baryon density.

\section*{Acknowledgments}
This work is partially supported by BMBF grant 05P12WOCTB and by DFG through CRC 110 ``Symmetries and Emergence of Structure in QCD". The authors would like to thank Yoichi Ikeda for 
providing them with his code for computing the $\overline{K}N$ scattering amplitudes. 





\end{document}